\documentclass[sigconf]{acmart}
\usepackage{enumitem}
\usepackage[ruled,linesnumbered]{algorithm2e}
\usepackage{mathrsfs}
\usepackage{makecell}
\usepackage{subcaption}
\usepackage{tablefootnote}
\usepackage{booktabs,tabularx}

\AtBeginDocument{%
  \providecommand\BibTeX{{%
    \normalfont B\kern-0.5em{\scshape i\kern-0.25em b}\kern-0.8em\TeX}}}

\copyrightyear{}
\acmConference{}
\acmBooktitle{}
\acmPrice{}
\acmDOI{}
\acmISBN{}

\newcommand{\LC}{\textsc{FasCo}}
\newcommand{\MSCN}{\textsc{MSCN}}
\newcommand{\DeepDB}{\textsc{DeepDB}}
\newcommand{\TLSTMCost}{\textsc{TLSTMCost}}
\newcommand{\QPPNet}{\textsc{QPPNet}}

\newcounter{siqiang}
\numberwithin{siqiang}{section}
    
\begin{document}

\title{Less is More: Towards Lightweight Cost Estimator \\for Database Systems}

\author{Weiping Yu}
\affiliation{%
  \institution{Nanyang Technological University}
  \country{Singapore}
}
\email{weiping001@e.ntu.edu.sg}

\author{Siqiang Luo}
\affiliation{%
  \institution{Nanyang Technological University}
  \country{Singapore}
}
\email{siqiang.luo@ntu.edu.sg}

\renewcommand{\shortauthors}{Weiping Yu and Siqiang Luo, et al.}

\begin{abstract}

We present {\LC}, a simple yet effective learning-based estimator for the cost of executing a database query plan. {\LC} uses significantly shorter training time and a lower inference cost than the state-of-the-art approaches, while achieving higher estimation accuracy. The effectiveness of {\LC} comes from embedding abundant explicit execution-plan-based features and incorporating a novel technique called cardinality calibration. Extensive experimental results show that {\LC} achieves orders of magnitude higher efficiency than the state-of-the-art methods: on the JOB-M benchmark dataset, it cuts off training plans by 98\%, reducing training time from more than two days to about eight minutes while entailing better accuracy. Furthermore, in dynamic environments, {\LC} can maintain satisfactory accuracy even without retraining, narrowing the gap between learning-based estimators and real systems.
\end{abstract}
\setcopyright{none}
\settopmatter{printacmref=false}




\maketitle

\let\clearpage\relax
\section{Introduction}
\label{sec:intro}



In database systems, estimating the running time of a query execution plan, abbreviated as {\it Cost Estimation}, is an indispensable function~\cite{selinger1989access}. Cost estimation is notoriously difficult, as the actual execution cost of a plan is highly dependent on a wide spectrum of factors including the involved operators, the underlying data distribution, and the resource availability. 

\begin{figure}[t!]
\centering
\vspace{-1mm}
  \includegraphics[width=0.7\linewidth]{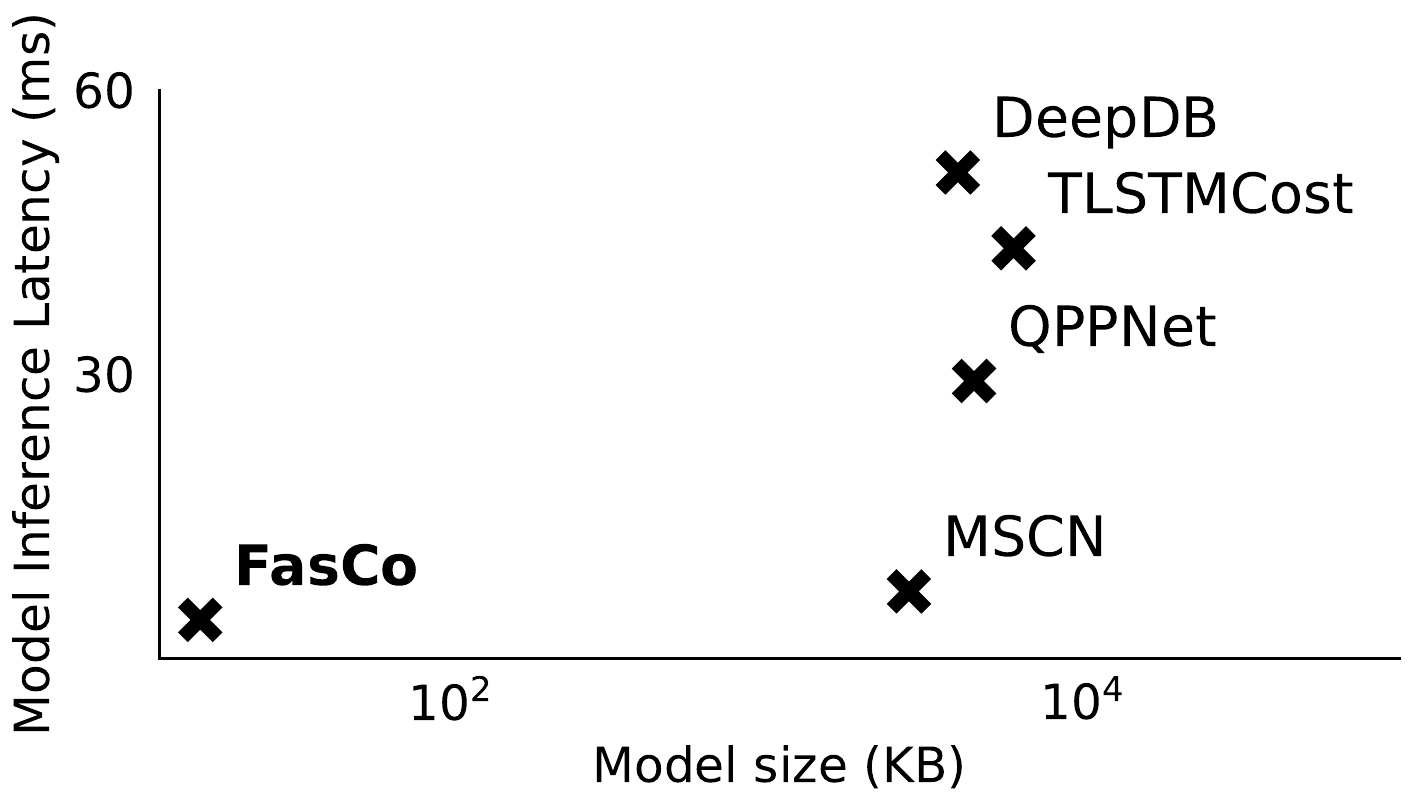}  
  \vspace{-4mm}
  \caption{Existing ML-based cost estimation models either
incurs large models or high latency (JOB-light workload).}
\label{fig:intro}
  \vspace{-6mm}

\end{figure}



Cost estimation has been extensively studied~\cite{leis2018query,lohman2014query, kipf2018learned,dutt2019selectivity,hasan2020deep,wu2021unified,yang2019deep,yang2020neurocard,hilprecht2019deepdb}. Classic cost estimators~\cite{leis2018query,lohman2014query} \textcolor{black}{embedded in the existing database systems often require a human expert to determine miscellaneous cost constants. 
} 
Recent researches~\cite{kipf2018learned,dutt2019selectivity,hasan2020deep,wu2021unified,yang2019deep,yang2020neurocard,hilprecht2019deepdb} employ advanced machine learning techniques to enhance the model expressiveness, aiming to relieve the burden of human experts and improve the estimation accuracy. Most models heavily rely on the estimation of the {\it cardinality}, which refers to the number of returned tuples after performing a certain query. When cardinality estimation is finished, the estimated execution cost is derived via an explicitly or implicitly learned relationship between the actual cost and the cardinality. These machine learning methods can achieve relatively high estimation accuracy~\cite{wang2021we,sun13end,marcus2019plan}.

Many existing machine learning-based cost estimators entail high model training or model inference costs, which are often rooted at the costly cardinality estimation process. Cardinality estimation approaches can be either query-driven~\cite{kipf2018learned,dutt2019selectivity,hasan2020deep} or data-driven~\cite{wu2021unified,yang2019deep,yang2020neurocard,hilprecht2019deepdb}. The former collects sufficient queries for model training, whereas the latter directly learns the cardinality from the data distribution. {Built upon these cardinality estimation models, the cost estimation models can be query-driven or data-driven, correspondingly.} While being effective, existing cost estimation models entail high training costs for large datasets. As an exemplar of a highly accurate query-driven cost estimator, {\TLSTMCost}~\cite{sun13end}, may necessitate multiple days of preprocessing or training for the benchmark JOB workload~\cite{leis2015good}, as pointed out by~\cite{hilprecht2022zero}. 
Data-driven approaches~\cite{hilprecht2022zero,hilprecht2019deepdb}, in contrast, require smaller training sets. However, they employ a relatively heavy model with a large parameter space for high estimation accuracy, leading to high training and model inference costs. For example, 
{\DeepDB}~\cite{hilprecht2019deepdb}, a cardinality estimator, has a model inference latency significantly higher than the vanilla estimator in PostgreSQL. Correspondingly, the cost estimator Zeroshot~\cite{hilprecht2022zero} built on top of {\DeepDB} has a longer model inference latency. 
The high training or inference cost can easily limit the scalability of these approaches, hindering their deployment to data-intensive applications.


\vspace{1mm}
\noindent
{\bf Our solution: \LC.} 
Our intuition is that the high cost of existing models either comes from training set size (e.g., a large number of execution plans in query-driven methods) or large model space (e.g., excessive parameters in data-driven methods). 
We argue that {\it a simple model can work better} in the context of cost estimation, motivated by which we design {\LC}, a \textbf{Fas}t \textbf{Co}st estimation model. Compared with state-of-the-art approaches, {\LC} exhibits a reduced model size, lower training overhead, and faster model inference latency, while still achieving superior or comparable accuracy with only about 2\% pre-generated training samples. As a preview, Figure~\ref{fig:intro} shows a comparison between {\LC} and the state-of-the-art approaches regarding the model size and model inference latency. {\LC}'s model size is at least two orders of magnitude smaller than {\DeepDB}, {\QPPNet}, and {\MSCN}, and it has an inference latency that is also about one order of magnitude shorter than {\TLSTMCost}. Interestingly, the design of {\LC} is surprisingly simple. Its model is a stack of multi-layer perceptrons (MLPs) following the execution plan tree. The effectiveness of the model is secured two-fold. First, {\LC} employs features that are selected carefully based on the internal structure of execution plans. Second, {\LC} incorporates a simple and effective data-driven cardinality calibration technique. 

\vspace{1mm}
\noindent{\bf Contributions.}
{Our contributions} are summarized as follows:

$\bullet$ One important contribution of this study is to build a way toward using lightweight cost estimation models for real-world database systems. Data are continuously updated in real-world database systems, and cost estimators have to be frequently retrained for high accuracy. Many existing methods focus on enhancing accuracy by integrating heavier machine learning models that require a high training cost. Inevitably, the existing high-cost models may be more difficult to be applied in such a dynamic environment. Our new model sheds light on that a lightweight machine learning model can be a better fit for this scenario. 

$\bullet$  We present {\LC}, a tree-structured cost estimator with cardinality calibration. Compared with previous tree-based models such as {\QPPNet}~\cite{marcus2019plan}, {\LC} reduces training costs and model inference latency significantly. {\LC} is simple for being only composed of small-sized MLPs, but as well effective for integrating many informative plan-related features to unleash the potential power of a simple model (see Section~\ref{sec:feature}). Furthermore, to compensate for the inaccuracy of cardinality provided by the histogram-based model, {\LC} employs a simple but very effective cardinality calibration method based on a sampling technique (see Section~\ref{sec:correction}).

$\bullet$  We conduct extensive experiments in real workloads on static and dynamic databases. Compared with existing works, both the accuracy and efficiency of {\LC} are significantly improved on various benchmarks. In addition, {\LC} also exhibits a significant advantage over the state-of-the-art methods when applied to dynamic database systems. 

\vspace{-1mm}
\section{FasCo}\label{sec:LC}

%


%
{\LC} is a model designed to accurately and efficiently estimate the runtime (cost) of an execution plan generated by a database system. The model consists of four main stages, which are outlined in Figure~\ref{fig:modeldesign}: First, {\LC} takes an execution plan corresponding to a SQL statement as input. The execution plan is a tree structure with nodes representing specific operations (e.g., {\it Sequential Scan}) required to execute the SQL statement. For each node, {\LC} then generates an embedding that conveys abundant information. The embedding of a parent node is generated by feeding the embeddings from its two children into a simple neural network. This allows {\LC} to be lightweight and efficient, using 98\% fewer training plans, incurring $40$ times lower training cost, and achieving higher accuracy than previous methods. While {\LC} incorporates a tree-structured network similar to QPPNet~\cite{marcus2019plan}, we highlight the main differences and novelties of {\LC} as follows. (1) \textbf{Efficient model design.} To accelerate training and inference, we propose a lightweight network architecture. First, we simplify the execution plan tree structure by merging each unary node and its only child node. As such, a unary node is no longer treated as an independent node and hence reduces the training and inference costs which are proportional to the number of nodes. Then, we share the information between the nodes effectively so that the model can be compressed to three sets of MLPs with only a few layers (as shown in Figure~\ref{fig:modeldesign}(d)). This allows us to avoid learning a separate model for each operator as in~\cite{marcus2019plan}. 
We find that such a simple architecture becomes surprisingly powerful when integrated with the features we carefully selected. Due to the simplicity of the model, this architecture significantly speeds up the training and inference. In addition, we incorporate a weighted Q-error into the loss function, making our model more accurate (see Section~\ref{sec:modeldesign}).
%
%
%
%
%
%
%
(2) \textbf{Cherry-picked explicit features.} { We integrate more features directly related to the execution plan as explicit inputs as shown in Figure~\ref{fig:modeldesign} (c). 
Previous works rarely rely on these explicit features (e.g., \textit{Subquery} and \textit{Cardinality}) for model training and some of them rely on adding hidden features to secure accuracy. We find that by incorporating these explicit features, we can significantly reduce the size of hidden features. We note that these explicit features are important to enhance the accuracy of the model. More details of feature extraction will be discussed in Section~\ref{sec:feature}.}
%
%
%
(3) \textbf{Cardinality calibration.} We also investigate the cardinality modeling given by histograms and propose {\it cardinality calibration} to further improve the accuracy.  This technique is based on sampling without sacrificing much efficiency.

\vspace{-2mm}
\subsection{Model Design}   
\label{sec:modeldesign}

\begin{figure*}[ht]
\vspace{-1.5mm}
  \centering
  \includegraphics[width=0.86\linewidth]{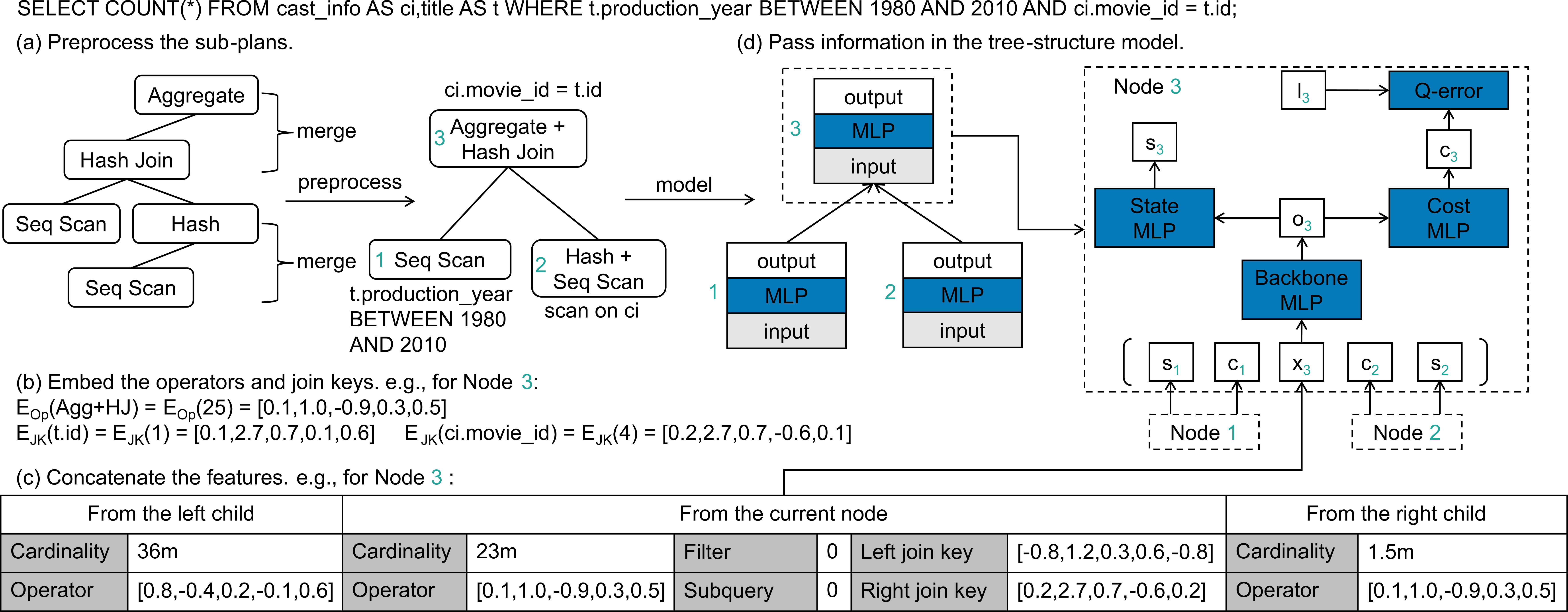}
  \vspace{-3mm}
  \caption{Model design of {\LC}.}
\label{fig:modeldesign}
\vspace{-5mm}
\end{figure*}

\noindent \textbf{Preprocessing.} The base structure of {\LC} is geared to the tree of the input execution plan. To improve the efficiency, we simplify the tree to have fewer nodes, making the subsequent operators simpler and more efficient. Particularly, among the nodes of all types of sub-plans, the unary node with a single child node contains less information than the other types of nodes. This is because most of its information comes from its immediate child node. For example, In Node 3 of the example tree in Figure~\ref{fig:modeldesign} (a), the {\it Aggregate} node computes the count of a set of input 
tuples from the {\it Hash Join} node. Hence, we merge each unary node to its child node as shown in Figure~\ref{fig:modeldesign} (a). By merging we mean to take the total cost of the two nodes as the cost of the merged node, and the merged node also contains other information that comes from the original child node. After merging, the unary node no longer acts as an independent node, and the whole structure is simplified, thereby reducing model inference latency and training cost. 

\noindent \textbf{Model architecture.} {\LC} adopts a tree-based model architecture because (1) it resembles the structure of the execution plan and hence gives a nice one-one correspondence between the network node and plan node; and (2) it was proved to be effective~\cite{marcus2019plan,sun13end,hilprecht2022zero}. Further, to make the model lightweight, we use MLP, which is the simplest neural network. Our experiments demonstrate that this simplicity does not lead to worse effectiveness. 

The training follows a post-order traversal of the execution tree, with each plan node corresponding to a model node that takes input from its two child nodes. For example, the three-node plan in Figure~\ref{fig:modeldesign}(a) (right) is transformed into a model with three corresponding components in Figure~\ref{fig:modeldesign}(d) (left).
The $j$-th and $k$-th nodes are the left and right child nodes of the current $i$-th node, and in this concrete example, $j$,$k$, and $i$ are 1,2 and 3 respectively. We collect three sets of information from them. The {\bf first one} is the feature $x_i$, which denotes the full information of the current node, and some information (e.g., cardinality) of its two child nodes. $x_i$ can be obtained in the execution plan tree and we defer the details to Section~\ref{sec:feature}. The {\bf second one} is the hidden states $s_j$ and $s_k$, which denote the implicit information derived from its left and right child nodes respectively. These hidden states help correct the cost and cardinality estimates and provide more information of context beyond just using the explicit features $x_i$. 
The {\bf third one} is the estimated costs $c_j$ and $c_k$ that are respectively from its left and right child node. As the ultimate goal is to get an accurate total estimated cost for the whole execution plan, the accumulation of the cost of each node is also crucial in establishing the final estimate.


The inference based on the model is performed in a propagation following the execution plan tree in a post-order manner. 
In Figure~\ref{fig:modeldesign}(d), the feature $x_3$ (for Node 3), together with ($s_1$, $c_1$) from Node 1 and ($s_2$, $c_2$) from Node 2, is input into an MLP (named Backbone MLP) to produce an output $o_3$, which is then fed into two MLPs (named State MLP and Cost MLP) to generate the state $s_3$ and cost $c_3$. These values are propagated upward along the tree similarly and finally, we get the estimated cost at the root node. 

The internal structure of these MLPs is designed to be concise with a stack of linear layers and $Tanh$ activation function. Meanwhile, the network size can be compacted to a few layers with only thousands of parameters, leading to extremely low latency.

\noindent \textbf{Loss Function.} Based on the aforementioned model and the corresponding inference, our goal is to minimize the estimate error at each node. The estimated cost $c_i$ for Node $i$ can be compared with the ground-truth cost $l_i$, giving us a loss as: $\text{Q-error}(c_{i},l_{i}) = max(\frac{c_{i}}{l_{i}},\frac{l_{i}}{c_{i}})$.
As each node may have different impacts on the final estimation, we add weights to aggregate the final loss as:
\vspace{-2mm}
\begin{equation}
\label{equ:3}
L=\frac{1}{n}\sum_i \lambda_i \text{Q-error}(c_{i},l_{i}),
\vspace{-1mm}
\end{equation}
where $n$ is the number of nodes in an execution plan and $\lambda_i$ is the weight of $i$-th node. 
%
%
%
 To determine how to set $\lambda$, we investigate the actual costs for different types of nodes. From a statistical perspective, we observe that the execution time of nodes using an index (index nodes, e.g., {\it Index Scan} and {\it Bitmap Scan}) are usually short and subject to fluctuation, which increases the difficulties in model learning. 
 On the contrary, the nodes without using an index (non-index nodes, e.g., {\it Seq Scan} and { \it Nested Loop}) would take longer, bringing more reliable cost information. Moreover, the index nodes' runtime is only a small fraction of the total execution time. Since our final target is to estimate the total cost, the estimation accuracy largely depends on the estimates at the non-index nodes. Hence, we assign larger $\lambda$ values for nodes without using indexes. The setting of $\lambda$'s will be further discussed in the experimental section.

\vspace{-3mm}
\subsection{Feature Extraction}   
\label{sec:feature}


This section introduces the details of feature $x_i$ for Node $i$. Unlike existing tree-based methods~\cite{sun13end,marcus2019plan} that employ excessive implicit features, resulting in significant training or inference overhead, we adopt more explicit input features to reduce complexity and improve the learning process. Our intuition is that involving more explicit plan-related features can significantly boost accuracy and efficiency. We will describe the selected features in detail.

\noindent \textit{\underline{Physical Operator}} is an influential factor in the execution process since the runtime of different types of operators varies widely.
We use an embedding layer to process this feature. The advantage is that the embedding layer avoids the dependency on the order of features when encoding. Instead, the neural network will automatically learn the relationship and distinction between them. 


\noindent \textit{\underline{Subquery}} in the context refers specifically to the relationship between sibling sub-plan nodes. If the current node is the second child of its parent and it is a {\it Subquery} of its sibling, it means that the node queries the data from the result of its sibling node. Otherwise, the node will query the data from a raw table. We simply encode the two conditions to ``1'' and ``0'' without embedding layers.

\noindent \textit{\underline{Cardinality}} associated in a node is the number of rows that are qualified as results after the execution of the operations of the sub-plan rooted at the node. We find that it is a vital factor in estimating the cost of an execution plan. 

The initialization of leaf node cardinalities is vital and can be divided into three cases: (1) When the current node is not a subquery of its sibling nodes, we set its first cardinality input that corresponds to the left child as the count of rows of the table it is querying, and the second that corresponds to the right child to 1; (2) When the current node is subquery of its sibling node and only has the filtering predicate, we set its first cardinality as the row count of the result tuples of its sibling node, and the second to 1;  (3) When the current node is a subquery of its sibling but has the join predicate, meaning it merges the results in its sibling node with a raw table, we set its left-child cardinality to the cardinality of its sibling node, and the right one to the row count of the table.



\noindent \textit{\underline{Filter}} is the number of conditions in predicates. To make it clear, in the example plan tree in Figure~\ref{fig:modeldesign}, the first, second, and third nodes have 2, 0, and 1 filters respectively. We find that it is useful for {\LC} to integrate this feature, because the query execution time is positively related to the number of filters used. 


\noindent \textit{\underline{Join Keys}} denote the attributes that are used when the tuples from two tables are joined. This feature conveys the information related to the execution cost because the join cost can be related to which keys to be joined between the tables. 
We also use an embedding layer to process this feature. For example, in the third node in Figure~\ref{fig:modeldesign} (a), the two join keys {\it t.id} and {\it ci.movie\_id} are converted to a vector. Note that the leaf nodes can also have join keys. Because some scan nodes are subqueries of their sibling nodes and have join predicates as mentioned in the feature \textit{Cardinality}. Similarly, some merge nodes may not have join keys because the join operation is done in one of their child nodes. In this case, we embed the join key as an extra category. 
\vspace{-4mm}
\subsection{Cardinality Calibration}
\label{sec:correction}
The cardinality of executing a sub-plan (i.e., a sub-tree in the plan tree) is an important feature for estimating the total plan execution cost. As the actual cardinality corresponding to each node in the plan tree is not known beforehand, an {\it estimated cardinality} is often used. Among the learning-based methods employing such an approach, there are two issues. First, the classic histogram-based cardinality estimator embedded in the common DBMS can be vulnerable, rendering the final estimate of the execution cost unreliable. Second, these methods often require a large number of training samples to fit the model. 
To alleviate these issues, we present a simple but effective cardinality calibration technique based on the classic histogram-based model. 

We first briefly introduce the histogram-based cardinality models adopted in common databases.
\textcolor{black}{The main idea is to first estimate the row count of each predicate by a histogram-based method based on several statistical techniques~\cite{postgresql1996postgresql}. Then the cardinality of a merge node can be estimated by the {\it product} of the cardinalities under its 
sub-predicates, and this propagates upward along the tree to estimate the cardinality of the root node of the whole plan. We note that the {\it product} is used based on the assumption that the predicates applied in different tables are independent. }
%
%



\begin{figure}[h]
\vspace{-4mm}
\centering
  \includegraphics[width=0.85\linewidth]{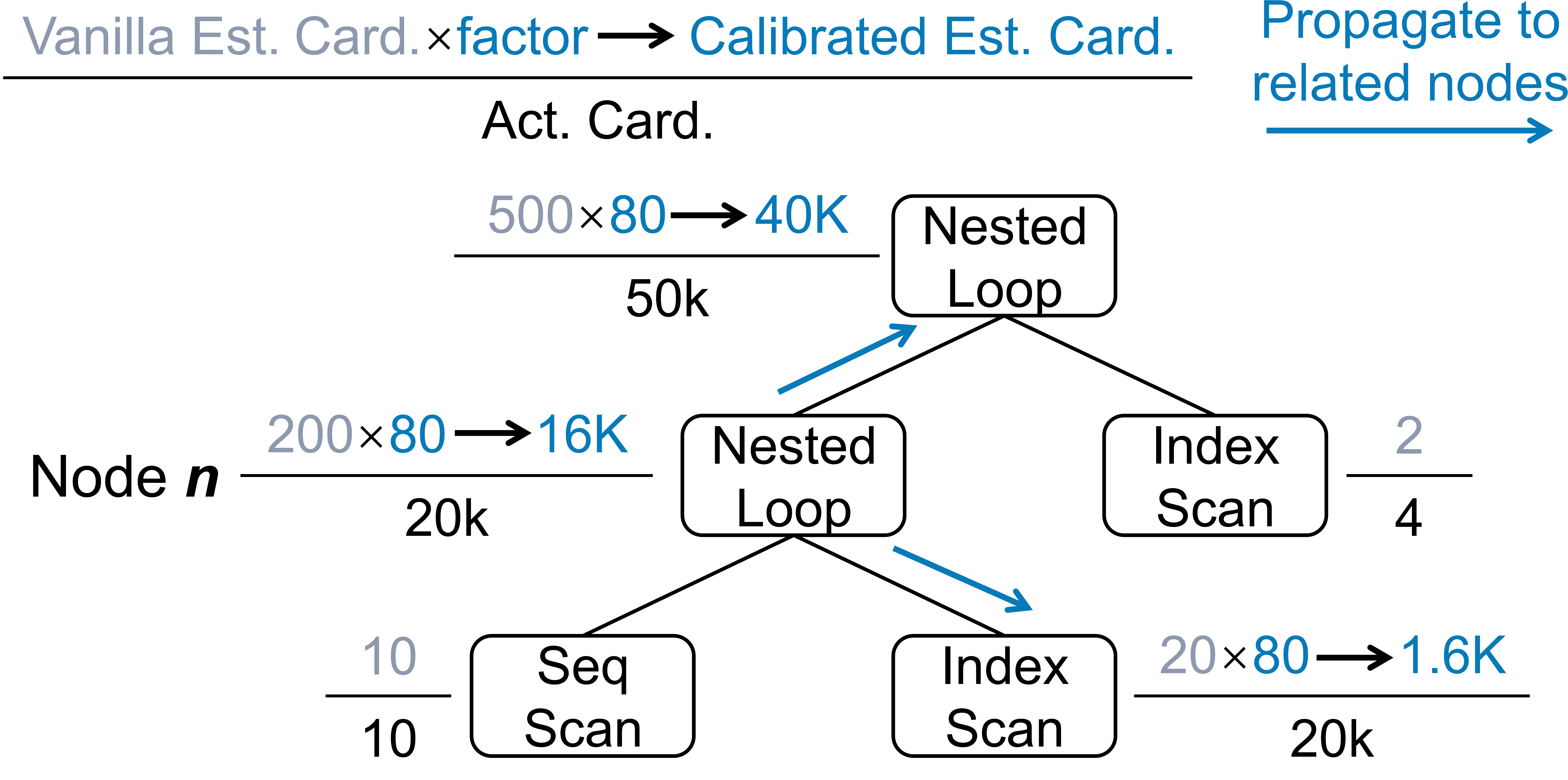}
  \vspace{-3mm}
  \caption{The process of cardinality calibration.}
\label{fig:correct}
\vspace{-4mm}
\end{figure}

The histogram-based method for cardinality estimation is known to be efficient as it requires no training. The downside is its low estimation accuracy for cardinalities. If we directly use this estimator for estimating the execution cost, we may suffer from low accuracy as well. 
{\it We argue that it is possible to apply a simple calibration of the classic cardinality estimator so that the cardinality estimation accuracy can be enhanced and becomes suitable to contribute positively to the cost estimator.} 

\begin{figure}[t]
  \centering
  \includegraphics[width=0.9\linewidth]{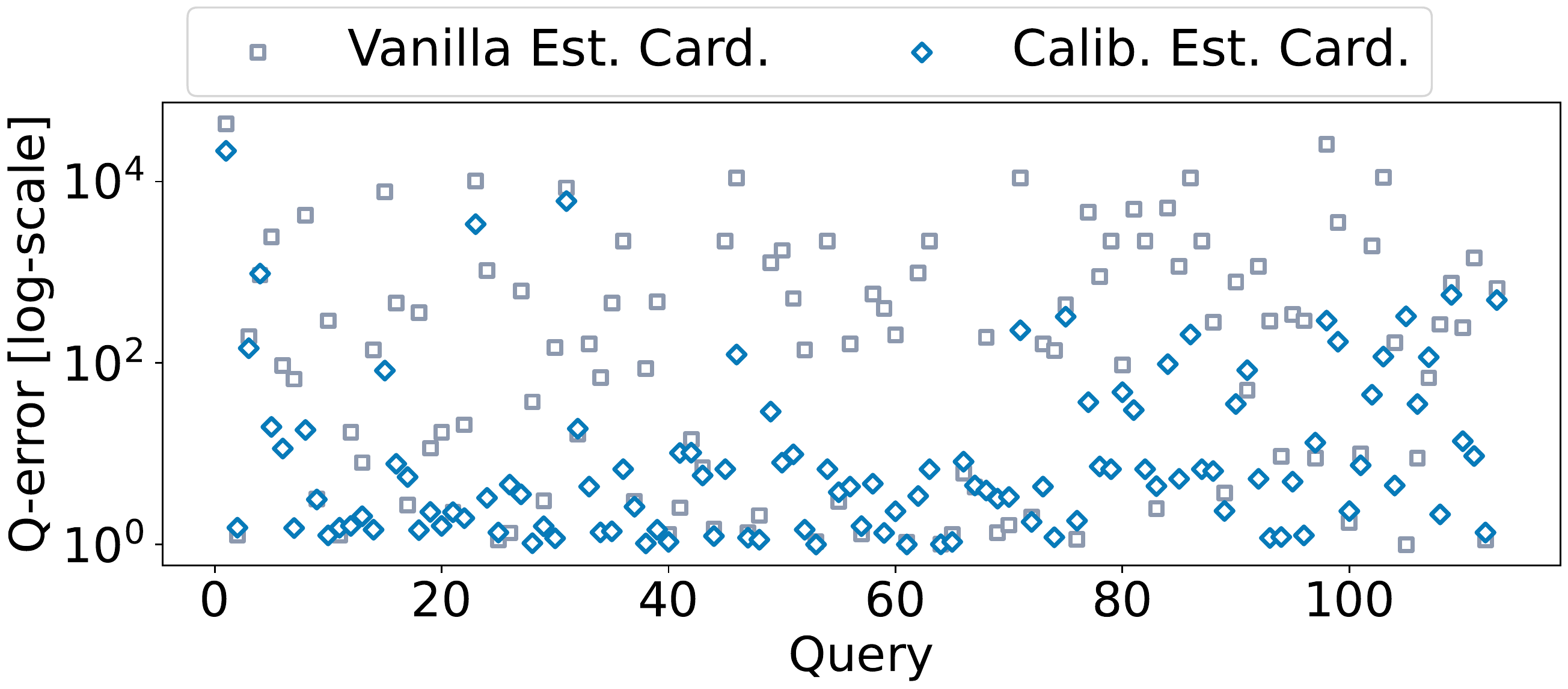}
  \vspace{-4mm}
  \caption{Q-error of vanilla estimated cardinalities vs. calibrated cardinalities on JOB-M workload.}
\label{fig:plan-actual}
\vspace{-6mm}
\end{figure}
Our main insight is that there is a strong correlation between the estimated cardinality of histogram-based methods and the actual cardinality. However, the accuracy of a histogram-based method is sometimes disrupted by disastrous estimates. As shown in Figure~\ref{fig:plan-actual}, we observe a significant gap between the estimated and actual cardinalities, but we do not that their cosine similarity is more than 0.90, which indicates a strong positive correlation between the two sets of cardinalities. 

The strong correlation indicates that the estimated cardinality can be calibrated. Serious errors that occur at merge nodes whose two children are both scan nodes often propagate to the remaining nodes. For example, Node {\it n} in Figure~\ref{fig:correct} is a merge node of this type, which are referred to as {\it lowest-level merge nodes}. The errors at {\it lowest-level merge nodes} propagate because the final estimate is the product of the cardinality estimates of all corresponding nodes. This also enlightens us that calibrating these rooted errors is more cost-effective. Therefore, we propose to {\it 
calibrate} the cardinality in these {\it lowest-level merge nodes} to avoid propagating the errors to the final estimate.
\noindent
{\bf Calibration Procedure.} Our method is based on sampling. First, in preprocessing we sample rows from the inner join results of each pair of tables with common attributes. The samples are stored in a {\it lookup list} and we set a low sample rate to store it in the main memory. Let us denote the sample rate as $\frac{1}{p}$ for some $p>1$. The sampling can be done in parallel to generate training plans. In {\LC}'s training and inference, we can query the corresponding lookup list to get a second cardinality estimate of the lowest-level merge node in the plan tree besides of vanilla estimate from DBMS. Particularly, we filter the samples in the lookup list with the predicate of the lowest-level merge node and get the qualified row counts, $c$. The second estimate for the merge node is then calculated as $c\cdot p$. If $\tilde{c}$ is the 
vanilla cardinality estimate, then the bias factor is $\frac{c\cdot p+{p}}{\tilde{c}+p}$.
%
%
In the numerator and denominator we both add a value $p$, which is the inverse of the sample rate. This helps counter the error that can be introduced when the sample rate is small, as it reduces the weight given to $c$.

Figure~\ref{fig:correct} illustrates the calibration process. In Node {\it n} there is a large estimate error (200 vs. 20k), causing an unacceptable error (500 vs. 50k) finally. To address the issue, during inference or training, our calibration method identifies Node {\it n} as a {\it lowest-level merge node} and inspects its lookup list. Then, we query the sampled lookup list and obtain a value of 239. Suppose that the sample rate is $\frac{1}{100}$, i.e., $p=100$. Then, the $factor$ is computed to be $\frac{239\times 100+{100}}{200+100}=80$. The calibration will be propagated to other nodes related to Node {\it n}.


To identify the related nodes of the Node {\it n}, the following conditions are used: First, all ancestor nodes of Node {\it n} are related to it, because their results come from it; Second, the sibling nodes of Node {\it n} and its ancestor nodes can be identified based on the {\it Subquery} (see Section~\ref{sec:feature}). If the node is the {\it Subquery} of Node {\it n} or its ancestors, it is relevant. Finally, the two child nodes of the Node {\it n} can also be distinguished by {\it Subquery} as the second condition. The estimated cardinalities of the related nodes are then calibrated by multiplying the factor, while the estimated cardinalities of unrelated nodes are left unchanged. This calibration improves the overall Q-error, as shown in Figure~\ref{fig:plan-actual}.

\noindent
\textcolor{black}{\textbf{Remarks.} Compared with fully sampling-based cardinality estimators, each lookup list in cardinality calibration only includes the merge results of two tables, which avoids huge memory overhead. Compared with other sampling strategies embedded in ML-based models, such as sampling bitmaps~\cite{wu2021unified,sun13end}, our cardinality calibration provides more precise and explicit information, which is more efficient for a lightweight model without extra embeddings of predicates.}

\section{EXPERIMENTS} 
\label{sec:experiments}


\begin{table*}[th!]%
  \caption{Accuracy comparison.}
  \vspace{-4mm}
  \label{tab:acc}
\resizebox{\textwidth}{!}{
\begin{tabular}{lcccccccccccccccccccc}\toprule
Workloads  & \multicolumn{4}{c}{JOB-M} & \multicolumn{4}{c}{JOB-light} & \multicolumn{4}{c}{JOB-light-ranges} & \multicolumn{4}{c}{TPC-H}& \multicolumn{4}{c}{Stack Overflow}   \\ \midrule
Methods    & Mean & 50th & 95th & 99th & Mean  & 50th  & 95th & 99th & Mean    & 50th   & 95th   & 99th   & Mean & 50th & 95th & 99th & Mean & 50th & 95th & 99th \\ \midrule
PostgreSQL &19.8&3.1&48.4&440.3&  9.5&2.8&31.1&125.7&         12.3&2.3&29.6&114.5&    3.0&2.7&6.7&7.9 &7.2&3.0&23.1&71.8     \\
MSCN       &   \multicolumn{1}{c}{--}   &    \multicolumn{1}{c}{--}  &   \multicolumn{1}{c}{--}   & \multicolumn{1}{c}{--} & 26.9&5.4&47.6&494.7&      31.8&11.4&467.6&494.7&   \multicolumn{1}{c}{--}     &  \multicolumn{1}{c}{--}    &   \multicolumn{1}{c}{--}     &  \multicolumn{1}{c}{--} & \multicolumn{1}{c}{--}& \multicolumn{1}{c}{--} & \multicolumn{1}{c}{--}  & \multicolumn{1}{c}{--}      \\
TLSTMCost  &6.9&4.0&17.6&64.7&   5.8&\textbf{1.9}&22.9&95.0&         6.2&2.5&30.9&88.0&   \multicolumn{1}{c}{--}   &    \multicolumn{1}{c}{--}    &    \multicolumn{1}{c}{--}& \multicolumn{1}{c}{--} & \multicolumn{1}{c}{--}  & \multicolumn{1}{c}{--} & \multicolumn{1}{c}{--}& \multicolumn{1}{c}{--}     \\
QPPNet     &8.1&4.4&20.9&79.6& 6.1&3.9&24.1&136.8&          8.1&4.7&37.1&99.3&  \textbf{1.2}&1.1 &\textbf{1.7}&1.8  &6.9&3.1&24.5&56.7    \\\midrule
\LC  &\textbf{4.3}&\textbf{2.4}&\textbf{13.6}&\textbf{41.1}&  \textbf{4.9}&2.0&\textbf{8.3}&\textbf{58.0}& \textbf{5.7}&\textbf{2.2}&\textbf{22.9}&\textbf{60.9}&   \textbf{1.2}& \textbf{1.0}&\textbf{1.7}&\textbf{1.7}&\textbf{4.2}&\textbf{2.1}&\textbf{11.6}&\textbf{39.7}     \\ \bottomrule
\end{tabular}}
\end{table*}

In this section, we evaluate {\LC} regarding (1) its accuracy and efficiency compared with the state-of-the-art cost estimators (2) its effectiveness in a dynamic environment (3) its ablation study.
\subsection{Experiment Setting}
\noindent \textbf{Workloads.} We conduct the evaluation on five workloads, namely JOB-M~\cite{leis2015good,yang2020neurocard}, JOB-light~\cite{yang2020neurocard},  JOB-light-ranges~\cite{yang2020neurocard}, TPC-H~\cite{poess2000new} and Stack~\cite{marcus2020bao}. The first three real-world workloads are derived from the JOB~\cite{leis2015good} workload that is coupled with the IMDB. The IMDB database contains 22 tables, and each table is joined by primary-foreign keys. We evaluate the family of JOB workloads on the test sets formatted in the previous researches~\cite{sun13end,wu2021unified,yang2020neurocard}. Following the training data generation in~\cite{sun13end}, we allow all possible values for operators and values without its restrictions. For TPC-H, we initialize a 100GB database and generate training and test plans by given templates. We also evaluate another real-world dataset, named Stack, which contains questions and answers from StackExchange websites, e.g., StackOverflow. Following~\cite{marcus2020bao}, we include 100GB of data and 25 query templates. For the last two workloads, we randomly split the 4000 template-generated samples into training and test sets in a 1:1 ratio.

\noindent \textbf{Metrics.} We use the {\it mean error} ({\it mean} for short), the {\it maximum error} ({\it max} for short), and the {\it $K$-th (quantile) error} ({\it $K$-th} for short) to measure the accuracy of the estimators. The error of each query is defined based on the most common Q-error~\cite{moerkotte2009preventing}. The mean and max are the averages and maximum errors while the $K$-th is the top-\{1-$K$\%\} largest errors in the tested workload.

\noindent \textbf{Baseline Methods.} 
We compare {\LC} to the following state-of-the-art methods: \textbf{{\MSCN}}~\cite{kipf2018learned}, \textbf{{\TLSTMCost}}~\cite{sun13end}, \textbf{{\QPPNet}}~\cite{marcus2019plan}, and \textbf{PostgreSQL}~\cite{postgresql1996postgresql}. We note that {\MSCN} does not support string predicates, and hence we only evaluate it on numeric workloads. Also, for \textbf{PostgreSQL} we need to transform the output by the EXPLAIN command to a time cost. We employ a simple linear transformation for this purpose and the transforming factor is selected by minimizing the mean Q-error of training sets.
\noindent \textbf{Environment.} We conduct all the experiments on Intel(R) Xeon(R) Gold 6326 CPU with 256GB Memory. The queries are executed with PostgreSQL 12.9 on Ubuntu 20.04.

\noindent \textbf{Hyper Parameters.} We utilize the Adam optimizer~\cite{kingma2015adam} and initialize the learning rate to 0.001. The model is trained for 10 epochs. The sampling rate of cardinality calibration keeps the size of each lookup list under 5MB, which will not impose a burden on memory. To demonstrate the generalization ability of our method, we use these settings on all the datasets and workloads.

\vspace{-4mm}
\subsection{Accuracy}

Table~\ref{tab:acc} presents the experimental results on all tested workloads. {\LC} achieves comparable or better accuracy against state-of-the-art estimators. The results show several major insights:

\noindent  \textbf{{\LC} consistently outperforms baselines in most cases.} For the JOB-M workload, {\LC} has achieved significantly lower errors than {\TLSTMCost} because {\LC} avoids the hardness of learning predicates embeddings. For the workloads of JOB-light and JOB-light-ranges with numeric predicates, {\LC} can still outperform {\TLSTMCost} and {\QPPNet} due to cardinality calibration. Particularly, the mean errors of {\LC} on JOB-light workload are 12\% and 16\% lower than those of {\TLSTMCost} and {\QPPNet}, respectively, and 8\% and 30\% lower for JOB-light-ranges. 
Although queries in Stack, another real-world dataset, are distinct from JOB, our method still consistently outperforms PostgreSQL and {\QPPNet}.

\noindent  \textbf{Transforming a cardinality estimator to a cost estimator is hard.} Interestingly, we find that a fine-tuned PostgreSQL can sometimes perform better than {\MSCN}, which is originally designed for cardinality estimation. For example, for JOB-light, PostgreSQL entails a mean error of 9.5, whereas {\MSCN} entails a higher mean error of 26.9. This result shows that even though {\MSCN} is well-performing in cardinality estimation, directly transforming it for the cost estimation can significantly degrade the performance. Particularly, {\MSCN} produces the cost in one iteration and the information about sub-plans is ignored. If a tree-structure estimator is to be implemented, {\MSCN} has to calculate the cardinalities of all the sub-plans, which will undoubtedly bring a great computational burden. In addition, {\MSCN} is hard to deal with string values in predicates, which is a common drawback of many query-driven methods. 



\noindent  \textbf{{\LC} has a comparable or better accuracy than {\TLSTMCost}.} {\TLSTMCost} achieves quite satisfactory accuracy in general, due to significant efforts have been paid in passing important information between nodes. In general, {\LC} has a comparable or better accuracy than {\TLSTMCost}. We note that there are two advantages of {\LC} over {\TLSTMCost}. First, original {\TLSTMCost} cannot handle some complex predicates in TPC-H such as predicates with the results from sub-queries, rendering it unable to produce results for TPC-H. In contrast, {\LC} can be applied to any type of predicates. Second, {\TLSTMCost} has to use much more training cost and processing time than {\LC} in order to achieve a comparable accuracy, as we will explain shortly in Section~\ref{sec:exp_efficiency}. 

\begin{figure}
  \centering
  \includegraphics[width=0.9\linewidth]{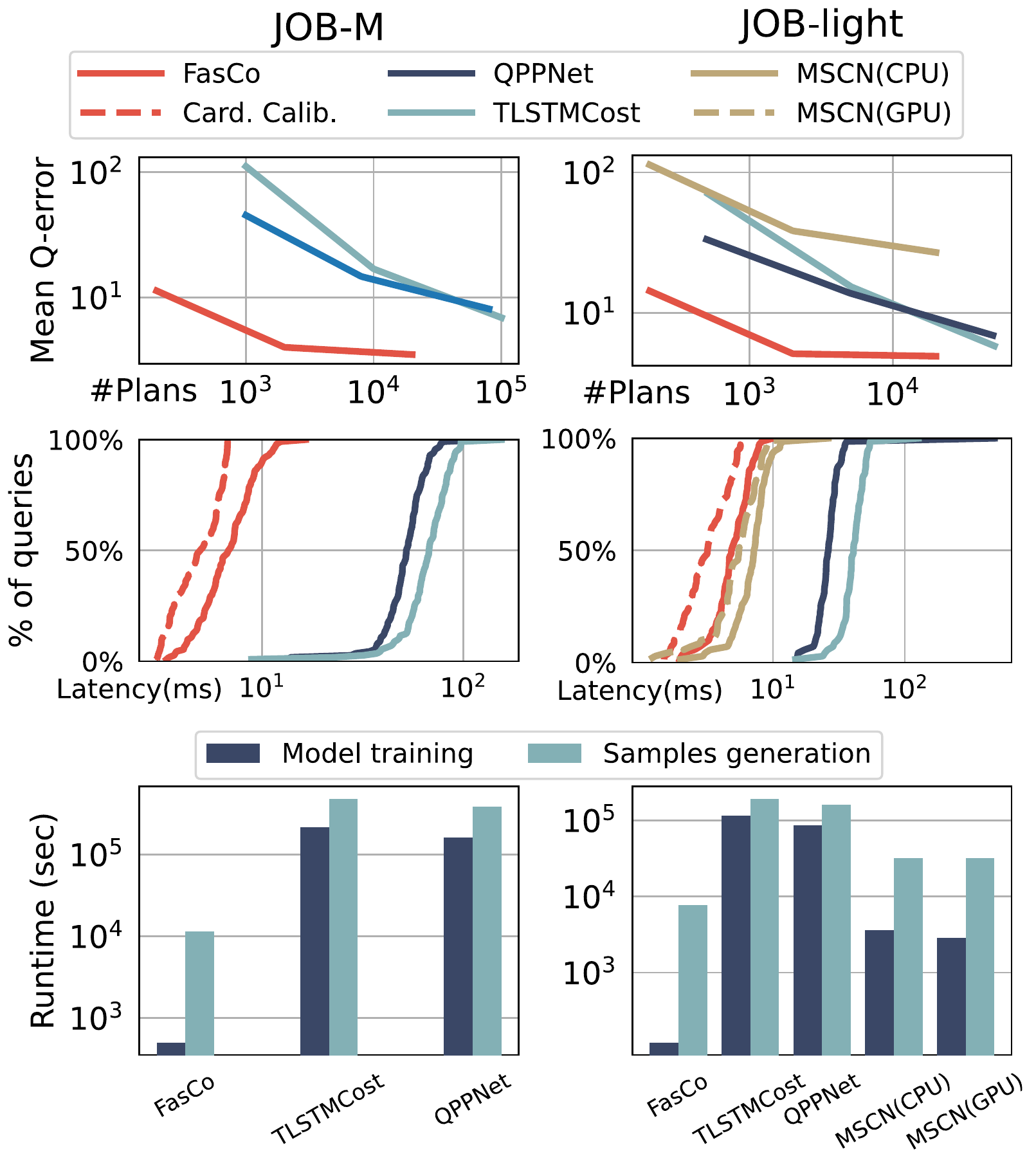}
  \vspace{-4mm}
  \caption{Efficiency comparison on JOB-M (left) and JOB-light (right) including accuracy vs. training plans, model inference latency and training cost. 
  }
\label{fig:eff}
\vspace{-6mm}
\end{figure}

\noindent  \textbf{{\LC} has a better accuracy than {\QPPNet} thanks to cardinality calibration.} Similar to {\LC}, {\QPPNet} also adopts a tree structure and statistical cardinality. However, due to the lack of an elaborated design, the model has to learn more implicit content (e.g. cardinalities of child nodes) on its own. In contrast, {\LC} cherry-picks more features as explicit input and adopts a more reasonable model design. In addition, because of the employment of cardinality calibration, {\LC} is better at handling cases with extreme errors, making its overall accuracy much better than {\QPPNet}.
\begin{figure}[th]
\vspace{-1mm}
  \centering
  \includegraphics[width=1.0\linewidth]{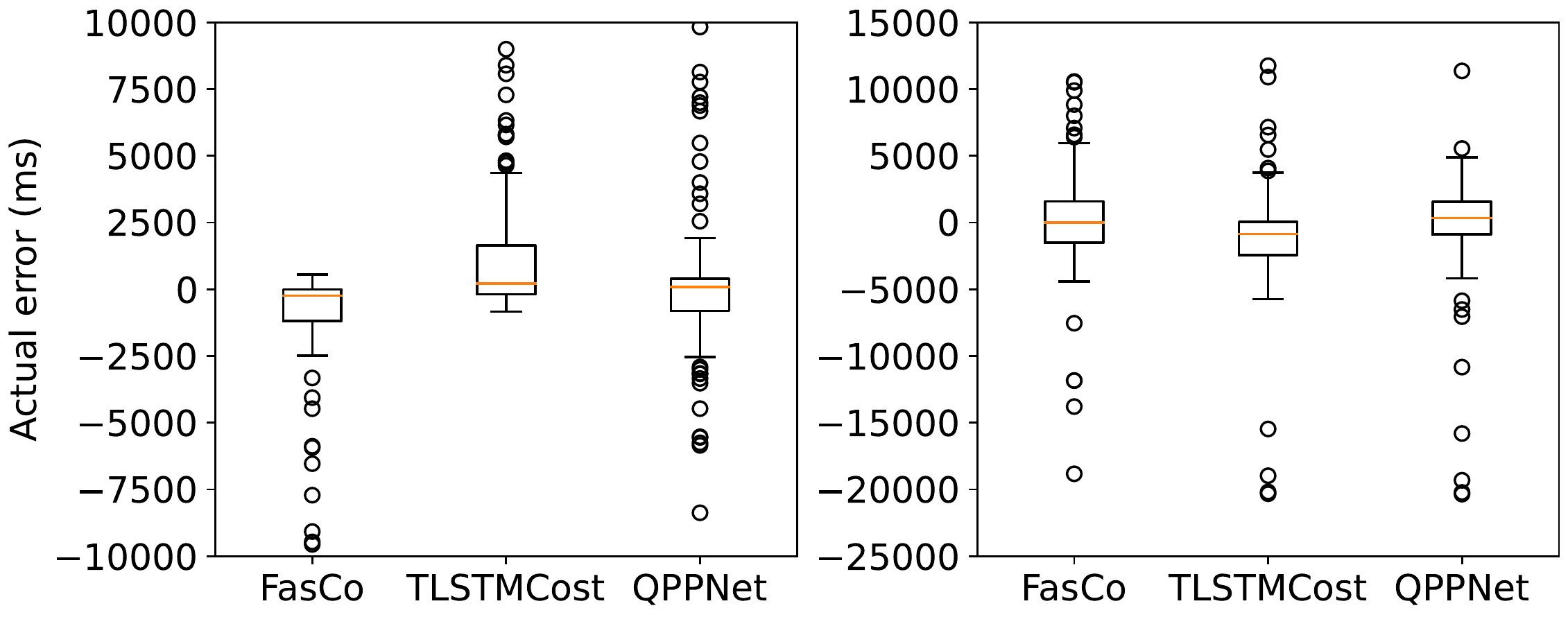}
  \vspace{-6mm}
  \caption{Actual errors on the JOB-M (left) and JOB-light (right) workloads. The box boundaries represent the 25th/50th/75th percentiles. Outliers are the points outside the box that are beyond 1.5 times the interquartile range.}
\label{fig:box}
\vspace{-4mm}
\end{figure}

\noindent  \textcolor{black}{\textbf{The actual error of {\LC} under different conditions is within a reasonable range.} In addition to Q-error, we also evaluate the actual errors (i.e. $cost_{est.}-cost_{act.}$ ) on JOB-M and JOB-light. As shown in Figure~\ref{fig:box}, although {\LC} sometimes may underestimate the cost of queries with abnormal statistical cardinalities, the cardinality calibration has significantly reduced the outliers compared to {\QPPNet}. Similarly, {\LC}'s outliers are less than the other two estimators on JOB-light workload with more skewed data.}
\subsection{Efficiency}\label{sec:exp_efficiency}


We compare the efficiency of methods by evaluating their accuracy on the JOB-M and JOB-light workloads. As shown in Figure~\ref{fig:eff}, {\LC} performs well with only 2000 training samples, while {\QPPNet} and {\TLSTMCost} require $80,000$ and $100,000$ training samples to achieve desired accuracy.

For sample generation and training, both {\QPPNet} and {\TLSTMCost} require multiple days to finish the whole process, whereas {\LC} takes only about 3 hours for samples generation and 8 minutes for training. {\LC} has a significantly improved efficiency as it uses only about 2\% samples of that in {\TLSTMCost}.




Inference latency is also mainly determined by the model size.  As shown in Figure~\ref{fig:eff}, {\LC} is about 7 and 10 times faster than {\QPPNet} and {\TLSTMCost} on average. Note that the latency of {\LC} includes cardinality calibration.

On JOB-light workload, {\LC} consistently outperforms {\QPPNet} and {\TLSTMCost} on model training, sample generation as well as model inference latency. {\MSCN}'s inference latency is close to {\LC} with GPU (GeForce GTX 1080 Ti) acceleration as shown in Figure~\ref{fig:eff}. 
From other perspectives, {\LC} significantly outperforms {\MSCN}.
The cost of the cardinality calibration is minor. It takes around 2-4 ms on average on the IMDB dataset with pre-loaded lookup lists, accounting for less than 50\% of the model inference latency.

In conclusion, due to the elaborate model design and efficient cardinality calibration, {\LC} can achieve better performance with fewer training samples and shorter latency.
\vspace{-1mm}
\subsection{Performance for Dynamic Scenarios}
We also conduct experiments to show that {\LC} is more suitable to be applied in a dynamic database environment, i.e., the data is gradually updating. This mimics the database application environment in the real world. Our setting is that the database does not need to collect training queries and retrain the model after relatively minor updates. 
We randomly delete 20\% tuples in each table except the type tables, such as {\it role\_type}. In this way, the 
correlation of the data is changed without loss of authenticity. Our update strategy for {\LC} is simple: we only update the lookup lists of cardinality calibration, which is described in detail in Section~\ref{sec:correction}. The cost of updating the lists is minuscule in comparison to the cost of collecting training data and retraining the model. For {\TLSTMCost} and 
{\QPPNet}, since we cannot find any such fast update strategy, we directly adopt the stale model on the updated database.

As shown in Table~\ref{tab:jobm-dacc}, the performances of methods relying on statistical cardinality, i.e. PostgreSQL, {\QPPNet} and {\LC}, are almost unchanged. Whereas the accuracy of {\TLSTMCost} drops significantly since the predicate embeddings still represent the original distribution. As the row counts, which are the dominant factor, have been re-analyzed, the updated histograms and cardinality calibration lead to satisfactory performance. 
\begin{table}\small %
  \caption{Dynamic accuracy performance on JOB-M.}
  \vspace{-3mm}
    \label{tab:jobm-dacc}
    \resizebox{0.95\linewidth}{!}{

  \begin{tabular}{lllllll}
    \toprule
    Methods & Mean & 50th & 90th & 95th & 99th & Max\\
    \midrule
    PostgreSQL & 39.9 & 4.27 & 89.9 & 212.3 & 322.4 & 456.5 \\
    {\TLSTMCost} & 9.9 & 7.0 & 29.2 & 40.6 & 124.7 & 247.1 \\
    {\QPPNet} & 8.2 & 4.6 & 20.1 & 23.2 & 88.0 & 101.0 \\
        \midrule
    {\LC} & \textbf{4.5} & \textbf{2.6} & \textbf{13.8} & \textbf{16.9} & \textbf{43.5} & \textbf{54.7} \\
  \bottomrule
\end{tabular}}
\vspace{-1mm}
\end{table}

Hence in practice, {\LC} can be adjusted to fit well in dynamic scenarios: first, we train the model with the initial database. Then, we
maintain lookup lists in the dynamic environment, and estimate costs by the stale model with updated cardinality calibration. When the degree of update reaches a user-defined threshold, we collect training samples and retrain the model. Then the new model can estimate the cost when its retraining is finished. Considering that other methods take a long time to collect training data and retrain the model, {\LC} can maintain a stable accuracy during this period.


\begin{table}\small %
  \caption{Ablation study of cardinality calibration (cc). }
  \vspace{-3mm}
  \label{tab:cc}
\resizebox{0.95\linewidth}{!}{

  \begin{tabular}{lllllll}
    \toprule
    Methods & Mean & 50th & 90th & 95th & 99th & Max\\
    \midrule
    {\LC} w/o cc & 7.3 & 4.3 & 10.7 & 18.3 & 77.7 & 89.2 \\
    {\LC} w/ cc & \textbf{4.3} & \textbf{2.4} & \textbf{8.2} & \textbf{13.6} & \textbf{41.1} & \textbf{48.5} \\
    {\LC} w/ bitmap & 6.7 & 4.2 & 9.2 & 18.0 & 62.1 & 79.5 \\
    {\QPPNet} w/ cc & 5.4 & 2.7 & 8.4 & 13.3 & 51.1 & 68.5 \\
    
  \bottomrule
\end{tabular}
}
\vspace{-1mm}
\end{table}

\begin{table}\small %
  \caption{Accuracy vs. $\lambda$. Notation $\mathcal{N}$ refers to the set of non-index nodes and $\mathcal{L}$ denotes the last node.}
    
  \vspace{-3mm}
  \label{tab:lam}
  \resizebox{0.95\linewidth}{!}{

  \begin{tabular}{llllllll}
    \toprule
    $\lambda$ of $\mathcal{N}$ & $\lambda$ of $\mathcal{L}$ & Mean & 50th & 90th & 95th & 99th & Max\\
    \midrule
    1 & 1  & 5.0 & 2.9 & 8.4 & 14.7 & 49.7 & 57.2 \\
    1 & 2  & 4.8 & 2.6 & 7.2 & 14.3 & 47.2 & 56.1 \\
    2 & 2  & 4.5 & 2.3 & 8.2 & \textbf{12.5} & \textbf{42.3} & 56.6 \\
    2 & 10  & 5.2 & 3.1 & 9.2 & 13.1 & 54.1 & 67.1 \\
    \midrule
    2 & 4 & \textbf{4.3} & \textbf{2.4} & \textbf{8.2} & 13.3 & 43.5 & \textbf{54.7} \\
  \bottomrule
\end{tabular}}
\vspace{-1mm}
\end{table}

\subsection{Dissecting {\LC} }
To gain more insights, we evaluate the importance of design components in {\LC} on JOB-M.

\noindent \textbf{Cardinality calibration.} It can be seen from Table~\ref{tab:cc} that the effectiveness of cardinality calibration is significant. The mean Q-error drops by 42\% compared to the method without calibration. The results show that if the cardinality calibration is removed, the performance of {\LC} is less accurate. Therefore, cardinality calibration significantly improves the accuracy of the method using statistical cardinality. When this technique is substituted by a sampling-based bitmap (a binary embedding following {\TLSTMCost}\cite{sun13end}), the improvements are compromised, since bitmaps provide less precise and explicit information than cardinality calibration. We also try embedding cardinality calibration into {\QPPNet} to prove its independent efficiency. The results show that though the accuracy cannot exceed {\LC}, the mean error has been reduced by 33.3\% compared with the original {\QPPNet} (5.4 vs. 8.1). However, the training cost is still much higher than {\LC} due to redundant parameters and less shared information between different plan nodes.


\noindent \textbf{Weights of Loss Function.}
In the loss calculation, the weight of each node, $\lambda$, is assigned based on the nodes' importance. So non-index nodes and the last node are given higher weights, while index nodes are assigned a weight of 1. As shown in Table~\ref{tab:lam}, when we appropriately increase the weights of the two types of nodes, the accuracy gradually improves.
The overall performance is optimal when the weights of the non-index nodes and the last node are 2 and 4. The average error is 20\% lower compared with the case when $\lambda$ is not adjusted. However, when the weights increase excessively, it will affect the learning in other nodes. This is why the accuracy drops drastically when the weight of the last node is 10. 

\begin{table}\small %
  \caption{Performance vs. network architectures.}
  \vspace{-1mm}
  \label{tab:arc}
  \resizebox{0.95\linewidth}{!}{
  \begin{tabular}{llllll}
    \toprule
      $bone$ &  $state$ &  $cost$ & Mean & 95th & Latency\\
    \midrule
    $1\times MLP$ & $1\times MLP$ & $1\times MLP$ & 4.5 & 14.7 & \textbf{7.2ms} \\
    $1\times MLP$ & $1\times MLP$ & $3\times MLP$ & \textbf{4.3} & 14.2 & 7.5ms \\
    $2\times MLP$ & $1\times MLP$ &$2\times MLP$ & 4.4 & \textbf{13.3} & 7.9ms \\
    $1\times MLP$ & $2\times MLP$ & $2\times MLP$ & 4.4 & 13.9 & 7.7ms \\
    $1\times CNN$ & $1\times CNN$ & $2\times CNN$ & 4.6 & 14.1 & 8.1ms \\
    \midrule
    $1\times MLP$ & $1\times MLP$ & $2\times MLP$ & \textbf{4.3} & \textbf{13.3} & 7.4ms \\
  \bottomrule
\end{tabular}}
\end{table}

\begin{figure}[]
\vspace{-1mm}
  \centering
  \includegraphics[width=0.9\linewidth]{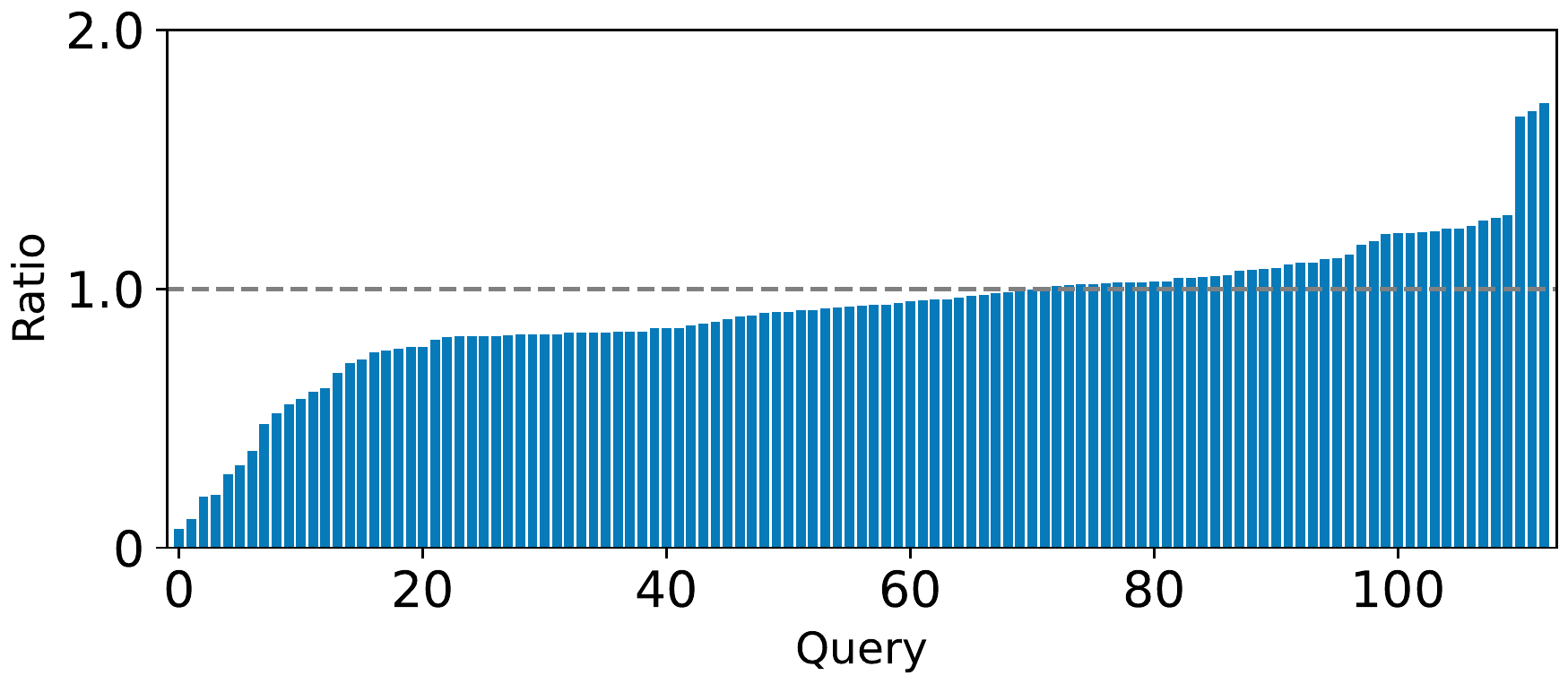}
  \vspace{-1mm}
  \caption{The ratio of JOB-M execution time before and after applying {\LC} to PostgreSQL. Lower ratio indicates better performance.}
\vspace{-1mm}
\label{fig:ratio}
\end{figure}

\noindent \textbf{Network Architecture.}
We conduct experiments on several combinations of the three sets of MLPs (see Section~\ref{sec:modeldesign}) and Table~\ref{tab:arc} shows the results. The default setting is 2 layers in Cost MLP and 1 layer in the other two MLPs. When the layers changes, the accuracy remains almost the same. This phenomenon demonstrates the accuracy of {\LC} is not very sensitive to the layers of MLPs. We also implemented a Convolutional Neural Network(CNN)-based tree according to~\cite{marcus2019neo,marcus2020bao,mou2016convolutional}. While the accuracy is not obviously improved, the latency and training time on CPUs are longer due to the slow sliding window operation.

\vspace{-1mm}
\subsection{Case Study: Embedded into PostgreSQL}
We apply {\LC} to PostgreSQL's plan optimizer. 
In a plan, we use {\LC} to estimate the cost of the lowest merge nodes and propagate the calibrated cardinalities and costs upwards. Each bar in Figure~\ref{fig:ratio} corresponds to a query, showing the execution cost ratio between the two plans selected by {\LC} and PostgreSQL. The ratios are sorted in an increasing order for better visualization. Compared with the default PostgreSQL, {\LC} performs better than PostgreSQL in around $80\%$ cases, with an average improvement of 7\% in execution time. We remark that {\LC} is independent of how the execution plans are formed. In PostgreSQL, the quality of the candidate plan worked out by its search method is relatively limited, which diminishes the benefit brought by {\LC}. In future works, we will experiment with stronger search methods employed by other systems such as Neo~\cite{marcus2019neo} and Bao~\cite{marcus2020bao}. We can foresee the potential benefits because of {\LC}'s low training overhead and demonstrated adaptability in dynamic scenarios.

\section{RELATED WORK}
\label{sec:related}
\noindent \textbf{Cardinality estimation.}
Histogram-based model is a representative data-driven method, which is also the most common method used in relational database management systems. 1-D histogram~\cite{selinger1979access} assumes all attributes are mutually independent and the selectivity of each column is built as a cumulative histogram. In addition, multidimensional histograms~\cite{poosala1996improved,poosala1997selectivity,gunopulos2005selectivity} are more precise than 1-D histograms by capturing inter-column correlations. However, these density models usually lose information between columns and tables by taking an independence or uniformity assumption. 
%
%
Recently, a fixed tree-structure density estimator called sum-product networks (SPN)~\cite{poon2011sum} has been applied to estimating cardinality. 
An SPN is a tree structure where each leaf node stands for an estimated selectivity of a subset on a row or a column split by the sum nodes or product nodes.  
Both~\cite{hilprecht2019deepdb} and~\cite{zhu2020flat} generalize SPN variants to cardinality estimation and achieve a superior performance on both accuracy and efficiency. 
%
Another approach to constructing a density cardinality model relies on deep autoregressive models from the machine learning community~\cite{germain2015made,nash2019autoregressive,radford2019language,vaswani2017attention}. The main idea is to learn the correlation between columns sequentially by the neural networks.~\cite{yang2020neurocard,yang2019deep} develop strategies to overcome the practical challenges in cardinality estimation, such as skipping the wildcards and factorizing complex columns. Unfortunately, these methods are still subject to representing predicates, and large models can also lead to excessive model inference latency.  

Query-driven cardinality estimators~\cite{markl2003leo,wu2018towards} leverage past or collected queries to predict cardinalities by building a mapping between them with regression models. Traditional machine learning tools such as KDE~\cite{heimel2015self,kiefer2017estimating} and mixture models~\cite{park2020quicksel} have been explored as hybrid methods in predicting row counts. Among these methods, {\MSCN}~\cite{kipf2018learned} is the representation of neural networks, which adopts a multi-set convolutional network to learn the correlations between joins.~\cite{dutt2019selectivity} applies a lightweight XGBoost on cardinality estimation, reducing model inference latency effectively. 




\noindent \textbf{Cost estimation.} 
There is a plethora of studies on cost estimation~\cite{sun13end,marcus2019plan,venkataraman2016ernest,wu2013predicting,zhang2005statistical,zhang2006xseed,wu2013towards,duggan2011performance,duggan2014contender,zhou2020query,hilprecht2022zero}. Some studies~\cite{zhang2005statistical,zhang2006xseed,wu2013towards,duggan2011performance,duggan2014contender,zhou2020query} belong to classic non-machine-learning-based methods. Among the machine-learning based approaches, Zeroshot~\cite{hilprecht2022zero}, {\TLSTMCost}~\cite{sun13end} and {\QPPNet}~\cite{marcus2019plan} are the state of the arts. Zeroshot~\cite{hilprecht2022zero} employs a cardinality estimator {\DeepDB}~\cite{hilprecht2019deepdb} to estimate the cost, and it needs many datasets to pre-train the model. In contrast, {\LC} does not require a pre-training phase. Our scope also does not include other pre-training works~\cite{paul2021database,lu2021pre}.

{\TLSTMCost}~\cite{sun13end} and {\QPPNet}~\cite{marcus2019plan} are the closest works to {\LC}, because they also employ plan-structured networks. Compared with {\TLSTMCost}~\cite{sun13end}, {\LC} uses a more lightweight model architecture. Also, in terms of performance, {\LC} demands much fewer training samples than {\TLSTMCost}, which reduces the training cost significantly. In addition, {\LC} does not need to learn the representation of predicates, which means it can be applied to any type of SQL query. In contrast, {\TLSTMCost} cannot be applied on complex SQL, such as some queries in TPC-H. \textcolor{black}{Recently there are some techniques that aim to improve the representations of predicates~\cite{zhao2022queryformer}. However, these embedding models are still plagued by complex predicates, such as the ``Like'' predicate, and they cannot effectively reduce the training cost on string predicates.}

Compared with {\QPPNet}~\cite{marcus2019plan} that first proposes the plan-structured network, the first and second differences as compared to {\TLSTMCost} also apply. Our {\LC} inherits the basic tree-based framework but our techniques including featurization and loss function along the training and inference process are surprisingly effective. Also, {\QPPNet} learns a model for each operator, which can lead to a long training time since the independent model cannot share information. In contrast, {\LC} learns all the operations in one model.  In addition, {\LC} adopts the cardinality calibration which significantly improves the accuracy while {\QPPNet} uses the original plan cardinalities.  
\textcolor{black}{
There are other problem settings around the cost model, such as the end-to-end optimizer. Generally, their cost models serve as components to estimate the costs of plans produced by certainly learned optimizers~\cite{marcus2020bao,marcus2019neo}. Our setting requires estimating the cost of an arbitrary plan. Therefore, these approaches are not suitable for our settings.}
\vspace{-2mm}
\section{Conclusion}
\label{sec:conclusion}
We introduce {\LC}, a novel lightweight tree-based neural network model designed to address the challenges of cost estimation. First, after comprehensive analysis, we adopt a lightweight tree-based model with more explicit inputs, which allows less training expense and model inference latency. Then, we present a fast sampling method to calibrate partial cardinalities in sub-plans, which significantly improves the accuracy of cost estimation. Experimental results demonstrate that {\LC} not only outperforms state-of-the-art solutions in accuracy, but also greatly improves the efficiency of training and inference. In a dynamic environment, {\LC} can be applied to 
minor updates without retraining, and the stale model can be replaced seamlessly with updated ones.



\bibliographystyle{ACM-Reference-Format}
\bibliography{sample-base}
\newpage
\mbox{}
\newpage

\end{document}